# Observation of first-order quantum phase transitions and ferromagnetism in twisted double bilayer graphene

Le Liu[1,2,†], Xin Lu[3,†], Yanbang Chu[1,2], Guang Yang[1,2], Yalong Yuan[1,2], Fanfan Wu[1,2], Yiru Ji[1,2], Jinpeng Tian[1,2], Kenji Watanabe[4], Takashi Taniguchi[5], Luojun Du[1,2], Dongxia Shi[1,2,6], Jianpeng Liu[3,7], Jie Shen[1,2,6], Li Lu[1,2,6], Wei Yang[1,2,6*] & Guangyu Zhang[1,2,6*]

[1] *Beijing National Laboratory for Condensed Matter Physics and Institute of Physics, Chinese Academy of Sciences, Beijing 100190, China*

[2] *School of Physical Sciences, University of Chinese Academy of Sciences, Beijing, 100190, China*

[3] *School of Physical Sciences and Technology, ShanghaiTech University, Shanghai 200031, China*

[4] *Research Center for Functional Materials, National Institute for Materials Science, 1-1 Namiki, Tsukuba 305-0044, Japan*

[5] *International Center for Materials Nanoarchitectonics, National Institute for Materials Science, 1-1 Namiki, Tsukuba 305-0044, Japan*

[6] *Songshan Lake Materials Laboratory, Dongguan 523808, China*

[7] *ShanghaiTech Laboratory for Topological Physics, ShanghaiTech University, Shanghai 200031, China*

† These authors contributed equally to this work.
\* Corresponding authors. Email: wei.yang@iphy.ac.cn; gyzhang@iphy.ac.cn

**ABSTRACT**
Twisted graphene multilayers are highly tunable flatband systems for developing new phases of matter. Thus far, while orbital ferromagnetism has been observed in valley polarized phases, the long-range orders of other correlated phases as well as the quantum phase transitions between different orders mostly remain unknown. Here, we report an observation of Coulomb interaction driven first-order quantum phase transitions and ferromagnetism in twisted double bilayer graphene (TDBG). At zero magnetic field, the transitions are revealed in a series of step-like abrupt resistance jumps with prominent hysteresis loop when either the displacement field ($D$) or the carrier density ($n$) is tuned across symmetry-breaking boundary near half filling, indicating a formation of ordered domains. It is worth noting that the good turnability and switching of these states gives a rise to a memory performance with a large on/off ratio. Moreover, when both spin and valley play the roles at finite magnetic field, we observe abundant first-order quantum phase transitions among normal metallic states from charge neutral point, orbital ferromagnetic states from quarter filling, and spin-polarized states from half filling. We interpret these first-order phase transitions in the picture of phase separations and spin domain percolations driven by multi-field tunable Coulomb interactions, in agreement with Lifshitz transition from Hartree-Fock calculations. The observed multi-filed tunable domain structure and its hysteresis resembles the characteristics of multiferroics, revealing intriguing magnetoelectric properties. Our result enriches the correlated phase diagram in TDBG for discovering novel exotic phases and quantum phase transitions, and it would benefit other twisted moiré systems as well.

# I. INTRODUCTION

Phase transitions are expected to occur at the instability when thermodynamic properties (density of states, susceptibility, or compressibility) or quasiparticle scattering rates diverge. In the strong coupling regime, i.e. $U/W \geq \sim 1$ where $U$ is the Coulomb repulsion energy and $W$ is the kinetic energy, a rich interplay of charge, spin, and orbital degrees of freedom contribute to versatile symmetry breaking phase diagrams where different correlated orders compete or coexist. Twisted graphene multilayers emerge as highly tunable flat band systems to realize these exotic phases, such as correlated insulators [1–9], superconductivity [10–17], and magnetism [7,8,11,18–22]. In these flat band systems, the presences of van Hove singularities (VHS) [3,9,12,23] and cascade transitions [24–26] contribute significantly to instabilities that lead to the symmetry breaking isospin polarizations and the resulted delicate ground states. Thus far, orbital ferromagnetism with magnetic domains have been observed in valley polarized phases [7,8,11,18–22] where Coulomb interaction and band topology play important roles. However, the long-range order of other correlated phases mostly remains unknown. In particular, the spin-related phenomena are rare and direct evidence of the magnetic order of the spin-polarized insulator has not been observed yet.

Here, we focus on the instability of phase diagram when spin fluctuations are critical in the vicinity of a spin-polarized correlated insulator in TDBG. With unique spin-polarized correlated insulators [2–5] and VHS-like phase boundary [3,23], TDBG is an excellent platform to study the spin-related phases and phase transitions. Importantly, by sweeping the carrier density $n$ or displacement field $D$ back and forth, we observe first-order phase transitions at zero magnetic field with a series of Barkhausen-like resistance jumps and electrical hysteresis. We further demonstrate the dominant role of the spin order by performing both in-plane and out-of-plane magneto transport measurements, and interpret these first-order phase transitions in the picture of correlation driven spin-domain percolations. In addition, we have observed abundant first-order phase transitions among normal metallic states from charge neutral point, orbital ferromagnetic states from quarter filling, and spin-polarized states from half filling.

# II. RESULTS

## A. *D*-field and Doping Tunable First-order Phase Transitions with Hysteresis at *B* = 0T

Figure 1(a) shows a typical phase diagram at $T = 30$ mK, i.e. a color mapping of longitudinal resistance $R_{xx}$ as a function of $n$ and $D$, for the TDBG device with a twisted angle of 1.35°, and Fig. 1(b) is the corresponding $R_{xy}$ at a small magnetic field. The mappings reveal a correlated insulator at half filling in the moiré conduction band ($v = 2$) surrounded by a halo-like structure, in consistence with previous works [2–5,23,27]. Such correlated states have been regarded as the signature of spontaneous symmetry breaking [3,23] that spin degeneracy is lifted inside the halo. The halo structure, being a phase boundary, separates two kinds of phases: the normal metal (NM) outside the halo, and the correlated states inside the halo. Accompanied by symmetry breaking, the band structure will be reconstructed by electron-electron interaction [28], leading to the spin-polarized metal (SP-M, at $v \neq 2$) or the spin-polarized correlated insulators (SP-CIs, at $v = 2$) as shown schematically in Fig. 1(d). When the Fermi level is tuned across the halo boundary at $v < 2$, the Fermi pocket changes from electron-type to hole-type with vanishing $R_{xy}$ at the boundary, depicted by the white contour in Fig.

1(b) and corresponding black contour in Fig. 1(a). These transport behaviors are related to the saddle point type van Hove singularities (VHS), at which the density of states (DOS) diverges. The topology of Fermi surface changes when the Fermi level crosses VHS, known as the Lifshitz transition. As shown in Fig. 1(c), the transport gap at $v = 2$ grows with $|D|$ from 0.2 to 0.5 V/nm slowly and then drops rapidly to zero within a range of 0.05 V/nm. The different trends on two sides suggest an abrupt phase transition between SP-CIs and normal metal at large $|D|$. These observations are coincident with our Hartree-Fock calculations which show a sharp decreasing correlated gap. All these results strongly suggest a first-order phase transition occurring between normal metal and SP-CIs (iii in Fig. 1(d)) at a large $|D|$.

Smoking gun evidence of the first-order phase transition at $B = 0$ T are revealed in the transfer curves when the gate voltages are swept back and forth in the white dashed box of Fig. 1(a). In Fig. 1(f) and 1(g), the longitudinal resistance $R_{xx}$ shows a hysteresis loop as $D$ or $v$ sweeps across the phase boundary in opposite directions. This loop is independent on sweep speed of gate voltage (Supplementary Note 2), and yet is sensitive to activating current (Supplementary Note 3) and temperature (Supplementary Note 6), demonstrating an intrinsic first-order phase transition between normal metal and SP-CIs. Besides, multiple jumps, most likely Barkhausen jumps [29], in hysteresis loops indicate the formation of orderly domains. Note that the abrupt transitions as well as the hysteresis at zero magnetic fields are well reproduced in a separate TDBG device D2 with a twist angle of 1.21° (Supplementary Note 9). Considering the spin polarization of the insulator, we conclude that there exist multiple spin-polarized ferromagnetic insulating domains in this regime. By taking the difference $\Delta R = |R_{xx}(+) - R_{xx}(-)|$ where +/- represents the sweep direction, we maps out the regime where first-order phase transitions occur, shown by the blue color in Fig. 1(e), along the halo boundary from $v = \sim 2$ to $v = \sim 1.7$. Notably, the transition regime is intimately related to the VHS in single-particle band structure calculated by continuum model (Bottom panel of Fig. 1(c)). The Fermi surface at $v = 2$ changes from annular with two pockets for medium interlayer potential difference $U_d$ (60 – 80 meV) into a simple surface with one single electron-type pocket for large $U_d > 80$ meV. This coincidence suggests the divergent DOS across Lifshitz transition plays a crucial role in the first-order phase transition, with the satisfied Stoner criteria $U*\text{DOS}(E_F)>1$ ($U$ stands for the correlation strength), inducing ferromagnetic instabilities of Fermi surfaces [30–33].

As an initial demonstration, we show that such stable and gate-tunable first-order transitions could be useful for memory. This is achieved by applying a pulse voltage on back gate in phase transition regime. As shown in Fig. 1(h), a pulse voltage as small as 20 mV could induce a transition between the low resistance state of few hundred ohms (Ω) and high resistance state of ~22 kΩ with highly tunable nature. The switching of such resistance states suggests that TDBG is a potential candidate for the memory devices working at cryogenic temperature.

**B. Ferromagnetic First-order Phase Transitions**

In addition to the $D$-driven and the doping-driven ones, we also observe magnetic field $B$-driven hysteresis. Let's first focus on the perpendicular magnetic field ($B_\perp$) dependence for a fixed $v = 1.95$ in Fig. 2(a), where the upper panel shows a color mapping of $R_{xx}(D, B_\perp)$ at $T = 100$ mK and the lower panel is a corresponding hysteresis $\Delta R(D, B_\perp)$. Clearly, decent non-zero $\Delta R$ exists only at a low magnetic field smaller than 1.5 T, and it follows the boundary that separating the metallic states with low resistances and the insulating states with high resistances in the phase diagram. Fig. 2(d) shows one representative $R(B)$ curve at $D$ = -0.527 V/nm, and it gives pronounced resistance hysteresis loop when the sweeping direction of $B_\perp$ is changed. The hysteresis includes multiple step-like transitions, indicating the formation of multiple ferromagnetic insulating domains, and the whole loop is mirror symmetric with respect to $B$ = 0 T. For a comparison, we perform similar measurements at parallel magnetic fields ($B_\parallel$) for a fixed $v$ = 1.95 in the same range of $D$, as shown in Fig. 2(b) and 2(e). The resulted phase diagram and the hysteresis loop at $B_\parallel$ are identical to those at $B_\perp$ < 1.5 T, suggesting that the magnetic first-order transitions and the hysteresis are due to spin degrees of freedom, instead of orbital.

The origin of spin order is further supported by the vanishing hysteresis and the reverse direction of phase boundary at $B_\perp$ >1.5 T, as well as the lower $R$ or $\Delta R$ at $B_\perp$ compared to those at $B_\parallel$, where the orbital Zeeman effect competes with the spin polarization [21,34]. The details of the isospin competition will be discussed later in the observation of other first-order phase transitions.

**C. Instabilities of the First-order Transitions**

The first-order transitions and the hysteresis are suppressed with the increase $T$. For instance, the $B$-driven hysteresis reduces as $T$ is increased and it totally disappears at $T$ = 3.5 K in Fig. 2(c) and 2(f); similarly, the $D$-driven hysteresis at $B$ = 0 T disappears at around $T$ = ~2.5 K (Supplementary Note 6). The suppression is also observed when the applied current ($I$) is increased (Supplementary Fig. 2(a-c)). Aside from the suppression of hysteresis, it needs a bigger critical field to realize the first-order transition when the $T$ or $I$ is increased. For instance, at $v$ = 1.95 and $D$ = -0.527 V/nm, the critical fields ($B_c$) of the $B$-driven first-order transitions are revealed in the $R(B_\perp, T)$ mapping in Fig. 2(f), where $B$ is swept from negative to positive and $T$ is swept from low to high temperature. Note that there are two critical fields, a smaller $B_{c1}$ (negative) for the insulator to metal transition and a bigger one $B_{c2}$ (positive) for the metal to insulator transition, indicated as the black and the red dots in Fig. 2(f), respectively. The critical field follow a power-law relation, i.e. $B_c - B_0 \sim T^\alpha$, where $B_0$ is the critical field at zero temperature limit and $\alpha$ represents the power law coefficient. The critical field corresponds to the minimum spin Zeeman energy $1/2 g_s u_B B_c$ for nucleation of ferromagnetic insulating domains, while the temperature represents the minimum thermal fluctuation $k_B T$ to break the ferromagnetic order. Here $g_s$ = 2 is the spin g factor, $u_B$ is the Bohr magneton and $k_B$ is the Boltzmann constant. We find that $B_0$ is around zero for $B_{c1}$ and a non-zero value for $B_{c2}$, and $\alpha$ is between 2 ~ 3 for both the $B_{c1}$ and the $B_{c2}$. It's notably that almost the same $\alpha$ is observed in another device D2 (Supplementary Note 9), suggesting a universal role in this critical phenomenon. Last but not least, the zero-field resistance $R_{xx}(B$ = 0 T) in Fig. 2(f) shows an insulating-like behavior at $T$ < 1 K, of ~ 6 kΩ, and then it suddenly drops below 300 Ω at $T$ = 1 K, showing a metallic behavior at $T$ > 1 K; alternatively, such an insulator to metal transition could also be achieved at $T$ < 1 K by applying a large current of 50 nA, as shown in Supplementary Note 2.

**D. Phase Separation and Percolations of Spin Ordered Domains**

The first-order transitions and the hysteresis, driven by either displacement, or carrier density, or magnetic field, are observed at metal insulator transition (MIT) [35] and could be interpreted within a picture of phase separation [36,37]. The free energy of first-order transitions could have multiple local minimum points in the order parameter space, with a stable state at zero and a metastable state at finite value (Supplementary Note 7). The instability of the phase diagram is driven by an interplay of $U/W$ (which can be effectively tuned by $D$ and $n$) and magnetic field, which favors the domain nucleation and growth, against thermal energy as well as the couplings between domains and the surrounding metallic electron sea, which melts the domains.

The $D$-field dependent first-order phase transitions are captured qualitatively in the above-mentioned model. Firstly, away from the phase boundary (Fig. 3(a)), it is a metallic state with $R < 3$ kΩ and no hysteresis. This agrees with a reduced $U/W$ away from the halo boundary, where negligible ferromagnetic insulating domains are surrounded by metallic electron sea, and thus the electrical conduction is dominated by metallic electrons. Then, as the $D$ approaches the phase boundary (Fig. 3(b)), it becomes a metastable insulating state with $R_{xx} \sim 10$ kΩ at $B = 0$ T, and multiple resistance jumps emerge in hysteresis as $B$ is changed. This is the case of large $U/W$, which might result in multiple domains densely distributed in real space at $B = 0$ T, and the magnetic field would boost the growth of domains (Fig. 3(c)). At the critical fields, some percolation paths suddenly disappear, which would contribute to the resistance jumps. Lastly, when $D$ is tuned inside the correlated insulating phase (Fig. 3(d)), i.e., $R_{xx} \sim 75$ kΩ, the hysteresis effect still exists but the resistance smoothly changes with magnetic field. The result suggests the absence of globally preferred spin orientation for different domains at zero magnetic field. The external magnetic field provides an anisotropic energy $E = 1/2 g_s u_B B$ by Zeeman effect and it tends to align the spin orientation from different domains, as shown in the bottom panel of Fig. 3(e). The presence of domains with different spin orientation might be due to the inhomogeneity of twisted angle and unexpected strain despite the most delicate sample fabrication [38–42]. A slight inhomogeneity in moiré systems alters the ground state on a microscopic scale, inducing a phase separation near the phase boundary.

**E. Competing Orders and Abundant First-Order Transitions**

Next, we discuss the competing orders and the resulted abundant first-order transitions in the phase diagram when the valley polarization starts to set in. Fig. 4(a) and 4(b) are $R_{xx}(n,B)$ and $R_{xy}(n,B)$ color mappings at $D = -0.51$ V/nm, respectively. Here, $R_{xx}$ and $R_{xy}$ are symmetrically and asymmetrically processed, respectively, in order to eliminate the crosstalk (see Supplementary Note 1). There are four different regimes in the phase diagram, and evident phase boundaries are reflected in the Hall resistance measurements. Fig. 4(c) shows three representative $R_{xy}(B)$ curves at different fillings, from which Hall carrier density ($v_H$) could be obtained by linear fitting with low filed data. At $v = 1.48$ (top panel), the linear fit by the black dashed line yields $v_H = 1.43$. The observation of $v_H \sim v$ indicates a four-fold degenerated Fermi surface adiabatically evolving from $v = 0$, a single particle picture described by the continuum model [43,44]. At $v = 2.3$, the Hall carrier density follows $v_H \sim v - 2$, agreeing with a two-fold degenerated Fermi surface developing from $v = 2$. The phase boundary between SP-CIs and SP-M keeps unchanged in the perpendicular magnetic fields, indicating that both states are spin-polarized. At $v = 1.8$ close to the SP-CIs (middle panel), the Hall response can be divided in two parts, ordinary Hall effect $v_H \sim v$ in low magnetic field (< 1.8 T), and additional anomalous Hall effect in high magnetic field (> 1.8 T). The observed Hall effect follows $R_{xy} = B/(ne) + R_{xy}^A$, where the first term corresponds to the ordinary Hall effect, and the second term $R_{xy}^A \sim M$ (magnetization) represents the anomalous Hall effect [45]. Generally, intrinsic anomalous Hall effect implies a large Berry curvature of the energy band. In TDBG, two valley-polarized subbands could carry opposite Chern numbers that are associated with the valley-contrasting orbital magnetism [21,43,44,46], and they are separated from each other with the increase of $B_\perp$ due to the orbital Zeeman effect. Consequently, a valley-polarized (VP) state with orbital ferromagnetism and a finite berry curvature becomes a ground state.

The competition of spin and valley is also revealed in the phase boundary between SP-CIs and VP states. At low fields $B_\perp < 1.35$ T, the phase boundary between SP-CIs and NM extends to lower doping levels with an increasing $B_\perp$, due to the dominating spin Zeeman effect. At $B_\perp > 1.35$ T, VP states emerge, and it gradually takes up most of the phase space with the increasing $B_\perp$. The observation suggests that VP states have lower energy than other ground states at high magnetic fields. In addition, while the phase boundary between NM and VP-M keeps extending nonlinearly to a lower carrier density with increasing $B$, that between VP-M and SP-CIs shifts to a higher density almost linearly. The different tendency indicates a stronger competition of spin and valley polarization near $v = 2$, where the spin-polarized states resist the invasion of valley polarization. Besides, the four different phases contribute a series of phase transitions. The phase transition between NM and VP-M (Fig. 4(d)) is ferromagnetic first-order phase transition, contributing abrupt resistance jumps and hysteresis. Those between VP-M and SP-CIs (Fig. 4(e)), as well as NM and SP-CIs (Fig.4(f)), are also first-order transition, evident from the hysteresis. Note that all first-order transitions occur near the halo boundary, accompanied by isospin competitions. The halo boundary, also being a phase boundary with diverging DOS, gives birth to the large isospin fluctuation that might be responsible for the first-order transition; by contrast, the transition between SP-CIs and SP-M is continuous and non-hysteretic (Fig. 4(g)) due to the same symmetry breaking induced by the spin polarization.

**F. First-order phase transitions at quarter filling**

The orbital Zeeman effect in high perpendicular magnetic field will induce fully isospin polarization near $v = 1$ and $v = 3$. As shown in Fig. 5(a), two new symmetry breaking Fermi surfaces emanates from quarter filling at $B_\perp = 2$ T. Here, the value of $v - v_H$ measures the deviation of the band filling before and after the symmetry breaking, and it indicates the degree of isospin polarization [47]. The state spreading from $v = 2$ to $v = 1$ along the halo boundary corresponds to the VP-M mentioned above. The linecut in Fig. 5(b) shows $v - v_H \sim 1$ (blue region), suggesting the VP-M is actually the incipience of the isospin-fully polarized state at $v = 1$. The series of Landau levels emanating from $v = 1$ in VP-M also suggest the emerging symmetry breaking Fermi surface (Supplementary Note 10). The other state takes up the area around $v = 3$ and experiences a correlated gap-like transition at $v = 3$. The linecut also shows $v - v_H \sim 3$ (violet region in Fig. 5(b)). These states at quarter filling are most likely the spin- and valley-polarized state (SVP) according to the theoretical calculations [34,48]. Furthermore, as shown in Fig. 5(c-d), the phase transitions between VP-M/SVP (quarter filling) and SP-M (half filling) with gate voltage driven hysteresis loops turn out to be the first-order phase transitions. The perpendicular magnetic field driven hysteresis loops for these two states also suggest the key role of orbital magnetization at quarter filling [11,22] (Fig. 5(e-f)).

## III. CONCLUSIONS AND OUTLOOKS

In TDBG, we have observed correlated first-order phase transitions between SP-CIs and normal metal at zero magnetic field, driven by carrier density and displacement field independently. The observations agree well with the scenario of Lifshitz transition as well as the rapidly decreasing energy gaps at the phase boundary from Hartree-Fock calculations. These observations are important that they unveil the long-standing mystery about the nature of the halo boundary and SP-CIs [3–5,23] by providing the smoking gun evidence for the spin-polarized ferromagnetism. We have also observed identical first-order metal-insulator transitions when either in-plane or out-of-plane magnetic field is applied, further demonstrating the existence of spin-polarized ferromagnetic domains. Moreover, we have observed abundant competing phases and accompanied first-order phase transitions between different ground states where valley degrees of freedom play a more important role.

Our observations suggest an instability with strong isospin fluctuations near the halo boundary at high displacement field, where the combined van Hove singularity and reduced band flatness leads to the strong Coulomb interaction effects. While the first-order transitions and the hysteresis could be captured within a picture of phase separation and percolations, more delicate theory and experiments are needed for a better understanding of this system. For instance, the first-order transitions and the hysteresis are highly tunable by electrical field and magnetic field, resembling multiferroics. The complicated flat bands in twisted multilayer systems, especially the reconstructed bands after Lifshitz transitions, might host both the electron and hole pockets; and the spatial separation of the electrons and holes at non-zero $D$ could lead to a possible formation of electrical dipoles. At some circumstances, it might eventually form a state where spin, charge, and the layer are locked, giving birth to multiferroics in twisted multilayers. The presence of both continuous phase transition and first-order phase transition near the halo boundary also deserve more investigations in TDBG as well as other twisted multilayers.

**ACKNOWLEDGMENTS**

We thank Kun Jiang, Zhida Song, Kam Tuen Law, Yu Ye, Guoqiang Yu, Erjia Guo, Shiliang Li for useful discussions. We acknowledge supports from the National Key Research and Development Program (Grant No. 2020YFA0309600), National Natural Science Foundation of China (NSFC, Grant Nos. 61888102, 11834017, 12074413), the Strategic Priority Research Program of CAS (Grant Nos. XDB30000000& XDB33000000) and the Key-Area Research and Development Program of Guangdong Province (Grant No. 2020B0101340001). J.S. acknowledge support from the Synergetic Extreme Condition User Facility at IOP, CAS. K.W. and T.T. acknowledge support from the Elemental Strategy Initiative conducted by the MEXT, Japan (Grant No. JPMXP0112101001), JSPS KAKENHI (Grant Nos. 19H05790, 20H00354 and 21H05233) and A3 Foresight by JSPS.

# Figure & Figure captions

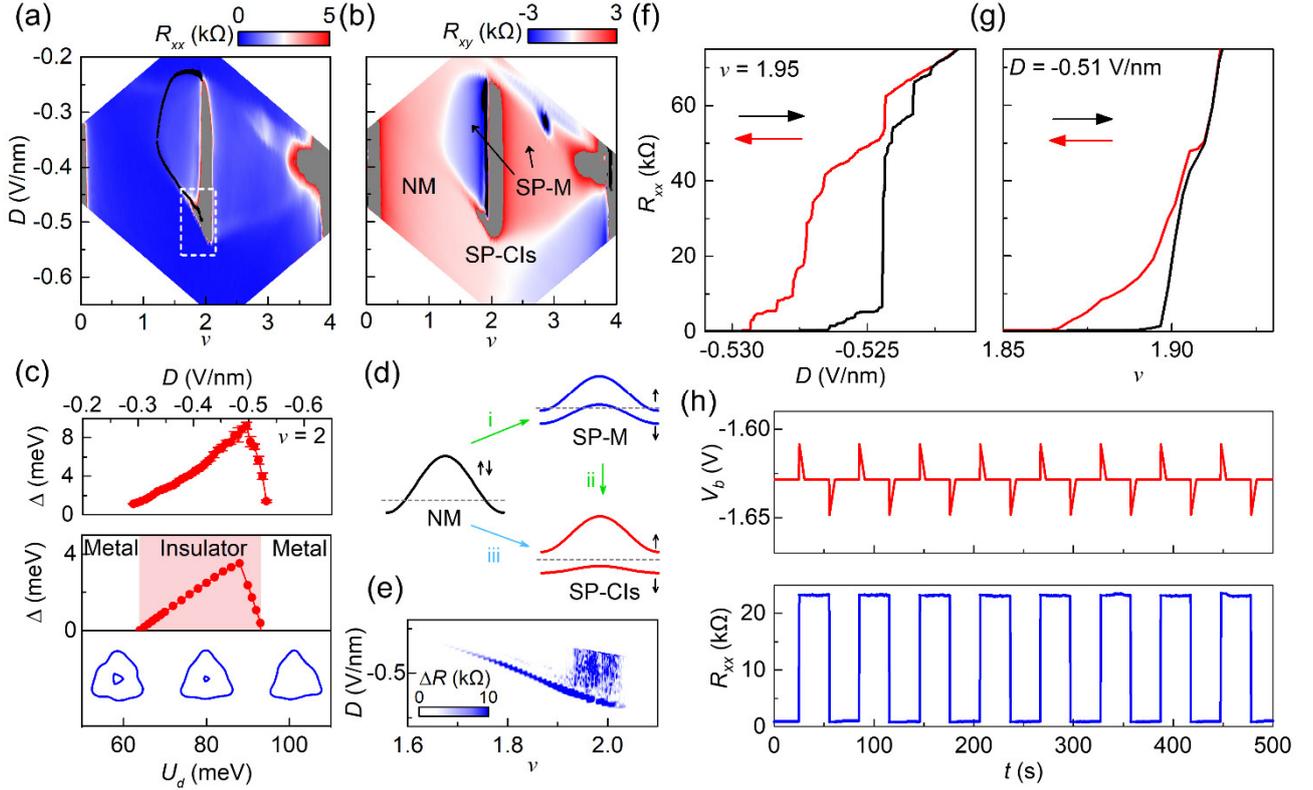

FIG. 1. Gate voltage driven first-order phase transitions at the halo boundary. (a), (b) $R_{xx}$ and $R_{xy}$ maps as a function of $v$ and $D$. The Hall resistance is measured at $B_\perp = 1$ T. The black dots in (a) correspond to the points of $R_{xy} = 0$ in (b). (c) Top panel: Thermal activation energy gaps versus $D$ at $v = 2$. Bottom panel: calculated correlated gap as a function of interlayer potential difference $U_d$ at $v = 2$. The corresponding Fermi surfaces are depicted in blue lines. (d) Schematics of different phases in (a) and (b). i, ii, iii correspond to phase transitions among three phases of normal metal (NM), spin-polarized correlated insulators (SP-CIs) and spin-polarized metal (SP-M). Black arrows correspond to directions of spin. Gray dashed lines correspond to Fermi levels. (e) $\Delta R$ maps limited in the white dashed box of (a). (f), (g) $D$ and doping driven Hysteresis loops. The black (red) line corresponds to the forward (backward) sweep direction. (h) Transitions between low and high resistance states by applying a sequential pulse voltage on the back gate.

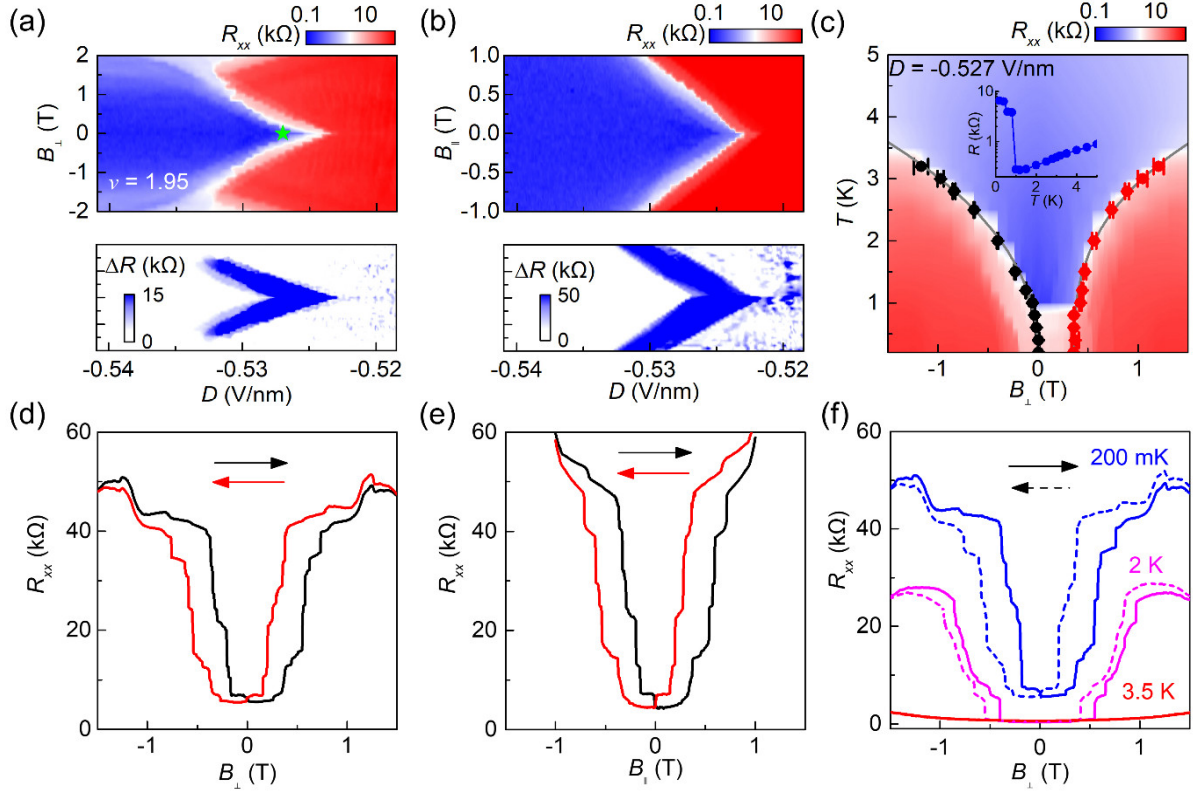

FIG. 2. Spin-polarized ferromagnetic insulators. (a) $R_{xx}$ and $\Delta R$ as a function of $D$ and $B_\perp$ at $v = 1.95$. (b) $R_{xx}$ and $\Delta R$ as a function of $D$ and $B_\parallel$ at $v = 1.95$. (c) $R_{xx}$ as a function of $B_\perp$ and $T$ at $v = 1.95$ and $D = -0.527$ V/nm. Critical magnetic fields $B_c$ are marked by black ($B_\perp < 0$) and red ($B_\perp > 0$) points. The gray line corresponds to the power-law fitting. Inset: $R_{xx}$ versus $T$ at $B_\perp = 0$. (d), (e) $B_\perp$ and $B_\parallel$ driven hysteresis loops at $D = -0.527$ V/nm and $T = 100$ mK marked by the green star in (a). (f) $B$-driven Hysteresis loops at different temperatures. The solid (dashed) line corresponds to the forward (backward) sweep direction.

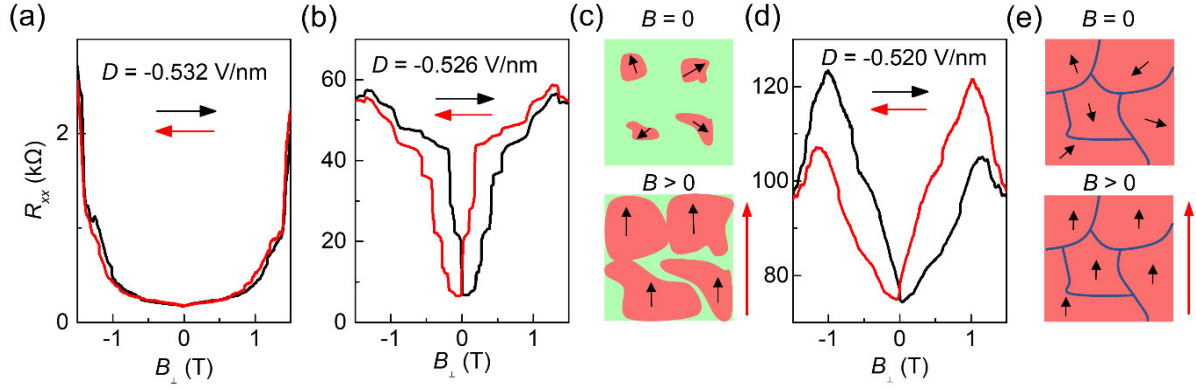

FIG. 3. Phase separation and percolation near the phase boundary. (a), (b), (d) $B_\perp$ driven hysteresis loops at different $D$ and $T = 100$ mK. (c), (e) Schematics of phase separation and percolation. Red regions correspond to spin-polarized insulating domains, where the black arrow is the orientation of spin polarization. Green regions correspond to normal metal. The red arrow is the direction of external magnetic field.

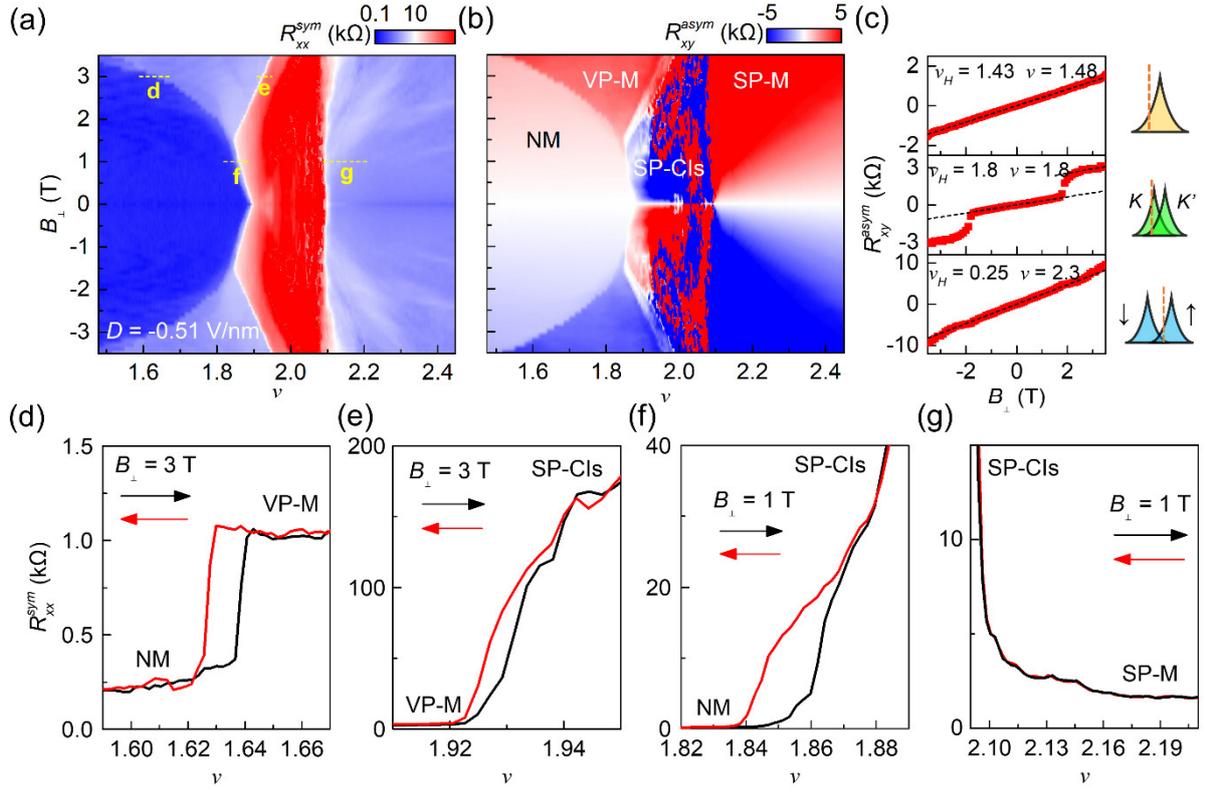

FIG. 4. Competing phase diagram and abundant first-order transitions. (a), (b) $R_{xx}$ and $R_{xy}$ maps as a function of $v$ and $B_\perp$ at $D$ = -0.51 V/nm. In figure (b), we label different phases as normal metal (NM), valley-polarized metal (VP-M), spin-polarized correlated insulators (SP-CIs), and spin-polarized metal (SP-M). (c) $R_{xy}$ versus $B_\perp$ at different filling factors. Hall carrier densities are extracted by linear fitting. Right figures are schematics of DOS in different phases. Top panel: schematic of DOS of four-fold degenerated bands. Middle panel: schematic of DOS of valley-polarized bands. Bottom panel: schematic of DOS of spin-polarized bands. (d)-(g) Doping driven Hysteresis loops of symmetrized $R_{xx}$ between different phases. All linecuts correspond to yellow dashed lines in (a).

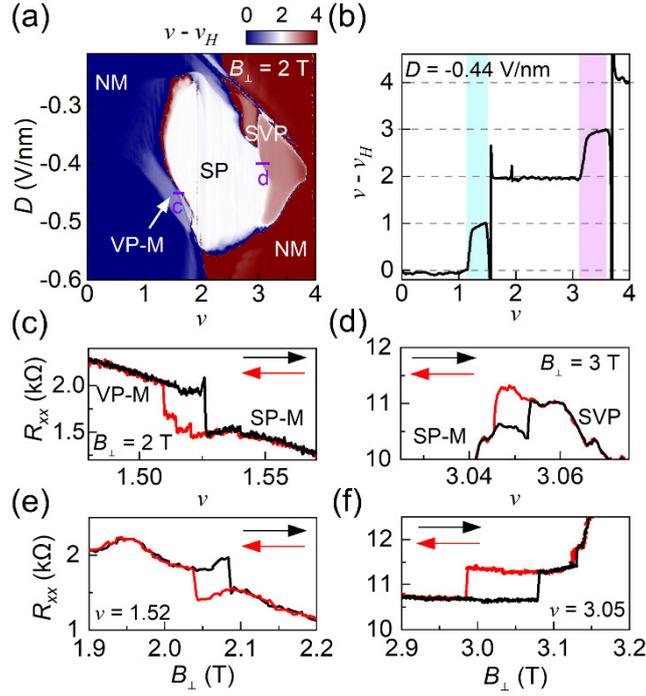

FIG. 5. First-order phase transitions and orbital magnetization at quarter fillings. (a) A color mapping of $v - v_H$ as a function of $v$ and $D$ at $B_\perp = 2$ T. Here, SVP corresponds to spin- and valley-polarized states. (b) A linecut at $D = -0.44$ V/nm from (a). The blue region corresponds to symmetry breaking states near $v = 1$, and the violet region corresponds to symmetry breaking states near $v = 3$. (c), (d) Typical doping driven hysteresis loop of $R_{xx}$ between VP-M and SP-M (c) and those between SP-M and SVP (d). The data in (c) and (d) correspond to the violet solid lines in (a). (e), (f) Magnetic field driven hysteresis loops of $R_{xx}$ at $v = 1.52$ and $3.05$, respectively.